# Efficient tensorized evaluation of permutation invariant polynomials for representing potential energy surfaces

Junhong Li,[1, #] Kaisheng Song,[1, #] Hua Guo,[2, *] and Jun Li[1, *]

[1] School of Chemistry and Chemical Engineering & Chongqing Key Laboratory of Chemical Theory and Mechanism, Chongqing University, Chongqing 401331, China

[2] Department of Chemistry and Chemical Biology, Center for Computational Chemistry, University of New Mexico, Albuquerque, New Mexico 87131, USA

[*] Corresponding authors, E-mails: hguo@unm.edu (HG), jli15@cqu.edu.cn (JL)

## TOC

## Abstract

Permutation invariant polynomials (PIPs), together with related polynomial-based invariant descriptors such as fundamental invariants (FIs), are widely used in constructing high-fidelity global potential energy surfaces (PESs) of molecules that contain identical atoms. Although the monomial symmetrization approach (MSA) enables fast evaluation of PIPs through recursive factorization, it leads to deeply nested computational graphs that may require large memory and are inefficient in modern automatic differentiation frameworks. In this work, we introduce JaxPIP, a JAX-based implementation that reformulates PIP/FI evaluation into tensorized linear algebra operations. By replacing recursive factorization with dense matrix operations combined with log-exp transformation and segmented summation, the evaluation becomes regular and GPU-friendly. This allows efficient execution with just-in-time compilation and enables large-scale batch evaluation of energies and forces (as well as higher-order derivatives). The resulting architecture supports ensemble simulations such as quasi-classical trajectory (QCT), path-integral molecular dynamics (PIMD), and diffusion Monte Carlo (DMC) in a fully vectorized manner. As demonstrated in examples, JaxPIP provides a practical route for efficient simulations of molecular systems with scalable GPU execution.

## 1. Introduction

Many modern machine learning potentials (MLPs) are built for local atomic environments.[1-7] The premise of such approaches is simple and based on intuition: most interatomic forces are "near-sighted", and this locality maps directly onto the neighbor-list scheme that molecular dynamics (MD) software has used for years. Global PESs can then be approximated by using this local approximation, for which the parameters are determined using ML tools such as neural networks (NNs). By balancing computational scaling, accuracy, and extensibility, this approach has become the mainstay for large-scale MD simulations with thousands of atoms or more. However, local MLPs may face challenges in accurately capturing non-local features or long-range interactions, because they rely on atom-centered cutoffs, even with sophisticated corrections such as message passing.[4, 5, 7, 8] Such problems become quite severe for reactive systems due to bond breaking and forming. The local approximation may also be insufficient to provide high accuracy necessary in spectroscopic calculations. Furthermore, despite their favorable scaling in large systems, local MLPs often carry a hefty computational pre-factor. This could become a heavy burden when a large number of force evaluations are required.

An alternative approach to MLPs is to construct global PESs with brute force. Such MLPs can be constructed with extremely high fidelity if sufficient data points are used. They are also expected to perform well for simulations involving configurations far from equilibrium, such as reaction asymptotes.[9, 10] This is why global-environment-based descriptors, such as internuclear distances,[11] continue to dominate high-precision modeling of small systems.[12-14] Because of the global nature of such PESs, it is essential to adapt the nuclear inversion and permutation symmetry,[15] not just for numerical efficiency, but also for a proper account of spectroscopic and dynamical attributes, particularly for floppy systems.[16, 17] To this end, the permutation invariant polynomials (PIPs)[18-21] and fundamental invariants (FIs)[22-25] have been proposed. (Note that the inversion symmetry is automatically satisfied because of the use of internal coordinates.) For example, many PESs based on linear combinations of PIPs (denoted below as the linear PIP method) have been reported by Bowman and coworkers,[12, 26] including systems with 21 atoms (e.g., Aspirin) via a single PIP basis[27] and reaching 44 atoms (e.g., *n*-$C_{14}H_{30}$) via fragment and many-body PIP approaches.[28, 29] In addition, Czakó and coworkers have developed many full-dimensional PIP PESs for polyatomic reaction dynamics beyond six atoms, especially for complex $S_N2$ and multi-channel reactions.[30-32] To improve the limited flexibility of the polynomial based expression, Jiang, Li, and Guo introduced PIP-NN, which frontloads NNs with PIPs to take advantage of their high expressiveness.[20, 21] As the number of atoms increases, however, the number of PIP terms grows combinatorially with both system size and polynomial order, leading to substantial numerical costs in evaluating of these PIPs. To mitigate this challenge, Zhang and co-workers introduced the FIs, which consist of the smallest set of PIPs that are sufficient to enforce the permutation symmetry.[22, 23] As a result, FI-NN significantly reduces the number of descriptors, pushing the boundaries of global PESs to much larger systems than before.[25] New FI generation algorithms are now capable of handling systems with ~20 atoms, such as naphthalene ($C_{10}H_8$) and Aspirin ($C_9H_8O_4$).[25]

It has been realized that the evaluation of the PIP descriptors could represent a major bottleneck in the global approach, especially for systems with large number of identical atoms. The current state-of-the-art is the monomial symmetrization approach (MSA).[18] MSA utilizes its recursive factorization scheme, which systematically identifies and reuses the common sub-expressions. By constructing high-order terms from computed low-order results, MSA effectively trades space for time to speed up. For derivative calculations, which are essential for MD simulations, MSA-2.0[19, 33] utilized scripts to automate the generation of PIP descriptors with analytical gradient code, but the workflow is still rooted in the Fortran era. MOLPIPx[34] represented an important step toward enabling native automatic differentiation (AD) for PIPs. As noted by the authors of Ref. [34], however, the direct translation may also encounter performance issues. For systems such as an $A_5$ molecule with $8^{th}$ order PIPs, for example, the just-in-time (JIT) compilation time exceeded 5 hours. This approach may also fail for even larger systems. This can be attributed to the inadequacy of the outdated computational structure based on Fortran compilers in modern AD engines.

Today, widely used frameworks like TensorFlow,[35, 36] PyTorch,[37, 38] and JAX[39] provide features like AD and low-level compiler optimizations that have become routine tools for modern ML development. For new generations of MLPs, obtaining analytical gradients or scaling to GPUs is considered as a standard expectation. The mathematical foundation for representing high-dimensional functions, such as PIPs, through structured multilinear algebra is well-established. As discussed by Debals and De Lathauwer,[40] evaluating multivariate polynomials is equivalent to performing contractions on high-order symmetric tensors. While MSA focuses on explicit algebraic factorization to reduce scalar arithmetic, a reformulation of the entire problem using a tensorized paradigm may significantly improve the performance, thus providing a marked advantage in numerical efficiency.

Here, we propose to shift our perspective from reducing arithmetic floating-point operations (FLOPs) to simplifying graph-tracing complexity within modern AD engines. To this end, we developed JaxPIP, which maps the permutation invariant basis into a dense tensor algebra, aiming to cater to the appetite of current AI frameworks and hardware accelerators, while retaining the accuracy requirements needed for high-fidelity simulations. The numerical savings are thus valuable for PIP based PES training and inference.

The rest of the manuscript is organized as follows: In **Section 2**, we demonstrate the architectural differences between MSA, MOLPIPx, and JaxPIP. Then in **Section 3**, we detail the software implementation of JaxPIP and benchmarks. The computational efficiency of JaxPIP with MSA-2.0 and MOLPIPx, and its applications to diffusion Monte Carlo simulations of glycine conformers and QCT simulations of the OH + $CH_3OH$ system are presented and discussed in **Section 4**. Finally, **Section 5** contains the conclusions and the outlook.

## 2. Methodology

### 2. 1 Factorized Evaluation via Monomial Symmetrization Approach

Pioneered by Xie and Bowman in 2010,[18] the MSA scheme has been the standard engine for evaluating PIPs for over 15 years. It was not only designed for automated generation of PIPs, but also for their optimized evaluation. As illustrated by the left pathway in **Figure 1(a)** and the detailed dependency graph, the core idea of MSA relies on a recursive bottom-up factorization scheme: higher-order polynomials are systematically decomposed into a basis of pre-computed lower-order monomials. Thus, redundant calculations of common intermediate terms are avoided, leading to a reduction in FLOPs. This strategy is closely related to common sub-expression elimination (CSE), a standard optimization strategy for reducing redundant computations.[41]

MOLPIPx[34] directly ported the forward evaluation of monomials and polynomials into JAX, aiming to seamlessly inherit analytical gradients via AD. This enables one to obtain analytical gradients and higher-order derivatives without manually deriving or generating significant symbolic codes. While the line-by-line recursive formulas generated by MSA-2.0 naturally align with the execution model of Fortran, graph-based AD frameworks such as JAX operate through tracing and optimization of computational graphs. To this end, the deeply nested scalar operations arising from recursive factorization of MSA may introduce long dependency chains, limiting opportunities for tensor-level optimization. As system size and/or polynomial order increase, the resulting graph complexity can substantially increase compilation overhead, as intermediate monomials must be first traced before subsequent optimizations are applied.

**2.2 Tensorized Evaluation in JaxPIP**

The fundamental building blocks of PIPs are the Morse-like variables (**y**), defined as transformed internuclear distances (**r**):[11]

$$y_k = \exp\left[-\frac{r_k}{\lambda}\right],\ k = 1, \cdots, N_d, \tag{1}$$

where $r_k$ is the $k$-th internuclear distance, $\lambda$ is the range parameter, $N_d = N(N-1)/2$ is the total number of distances for an $N$-atom system.

Following the original convention in the MSA,[18] a general primitive monomial is defined as a fully expanded product of Morse-like variables $y_k$ raised to specific integer powers. To distinguish these from the factorized intermediate terms, we denote them as flat monomials ($\boldsymbol{\mu}$):

$$\mu_i = \prod_{k=1}^{N_d} y_k^{B_{i,k}}, \tag{2}$$

where $B_{i,k}$ represents the basis element that represents the exponent of the $k$-th distance variable in the $i$-th flat monomial. The exponent sequence ($B_{i,1}$, $B_{i,2}$, …, $B_{i,Nd}$) corresponds to the shorthand notation [*abcdef*…] introduced by Xie and Bowman.[18]

Evaluating $\mu_i$ requires computing the exponentiation of $y_k$ for each distance, followed by a sequence of scalar multiplications. To rewrite the evaluation in a form compatible with dense tensor operations, we employ an identity transformation by mapping the product into the log-space and then exponentiating the results:

$$\mu_i = \exp\left[\ln\left[\prod_{k=1}^{N_d} y_k^{B_{i,k}}\right]\right]. \tag{3}$$

Using the property of logarithms, the product transforms into a summation:

$$\mu_i = \exp\left[\sum_{k=1}^{N_d} B_{i.k} \cdot \ln(y_k)\right]. \tag{4}$$

Substituting the definition of $y_k = \exp\left[-\frac{r_k}{\lambda}\right]$ into the logarithmic term, we can observe:

$$\ln(y_k) = \ln\left[\exp\left[-\frac{r_k}{\lambda}\right]\right] = -\frac{r_k}{\lambda}. \tag{5}$$

Consequently, the explicit computation of logarithms is eliminated. The expression for the $i$-th monomial simplifies to

$$\mu_i = \exp\left[\sum_{k=1}^{N_d} B_{i,k} \cdot (-\frac{r_k}{\lambda})\right]. \tag{6}$$

This scalar summation forms the inner product between the $i$-th row of the basis matrix and the distance vector. By vertically stacking the shorthand exponents [*abcdef*…] of all flat monomials, we construct an $N_\mu \times N_d$ dense exponent matrix ($\mathbf{B}$). The evaluation of the full flat monomial vector ($\boldsymbol{\mu}$) can therefore be expressed as a matrix-vector multiplication followed by an element-wise exponential:

$$\boldsymbol{\mu} = \exp\left[-\frac{1}{\lambda}\mathbf{B}\cdot\mathbf{r}\right]. \tag{7}$$

The same formulation can also be extended to alternative kernels commonly used in FIs. For the reciprocal kernel $y_k = 1/r_k$, for example, we have $\ln(y_k) = -\ln(r_k)$. Thus, the corresponding expression for FIs would become:

$$\boldsymbol{\mu} = \exp[-\mathbf{B}\cdot\ln\mathbf{r}] = \mathbf{r}^{-\mathbf{B}}. \tag{8}$$

Finally, to recover the polynomials ($\mathbf{p}$), the flat monomials ($\boldsymbol{\mu}$) are aggregated via segmented summation. This reduction is guided by a segmentation vector ($\mathbf{s}$) pre-calculated during the basis generation. It maps each flat monomial ($\mu_i$) to its corresponding symmetrized polynomial ($p_j$).

$$\mathbf{p} = \mathrm{SegmentSum}(\boldsymbol{\mu}, \mathbf{s}). \tag{9}$$

It is worth noting that this tensorized formulation is not limited to PIPs. FIs, which also serve as a descriptor for invariant PESs, share the same underlying structure of symmetrized monomials. Thus, this formulation provides a unified implementation pathway for polynomial-based invariant descriptors. Importantly, this tensorized formulation transforms the irregular, sparse dependency tree of factorization into a contiguous, dense linear algebra operation, ideally suited for modern AD frameworks and accelerator-oriented execution models.

**2.3 Implementation and Execution Workflow**

JaxPIP is implemented on top of the JAX framework, with Equinox used as a lightweight PyTree-based module interface. The descriptor is fully expressed in JAX primitives, enabling AD and JIT compilation without manual gradient derivation. This implementation directly realizes the tensorized formulation described in **Sec. 2.2**.

The computational workflow follows a straightforward pipeline. Atom-pair indices are first generated via `jnp.triu_indices`, followed by computation of internuclear distances. These distances are evaluated using the tensorized formulation in **Eq. (7)**, implemented as `jnp.exp(jnp.dot(B, r) * (-1.0 / _lambda))`. The resulting flat monomial vector is aggregated into PIPs via segmented summation as described in **Eq. (9)**. Force evaluation is obtained using `jax.value_and_grad`, ensuring strict consistency between energies and forces. Also, the formulation of JaxPIP is numerically stable in practice, as the exponential argument is bound above by zero, preventing numerical overflow across a wide range of molecular geometries. Additionally, since the orders of PIPs or FIs are non-negative and relatively small integers, the basis matrix **B** can be stored using the Uint8 data type, which reduces memory consumption to 1/8 of that required by standard FP64 data type, while remaining absolute numerical precision.

The differentiation in JaxPIP is based on reverse-mode AD implemented in JAX/XLA. Unlike symbolic or code-generated gradient schemes used in MSA-based Fortran implementations, the computational cost of gradient evaluation scales proportionally with the forward evaluation graph, without introducing additional symbolic expansion of intermediate monomials. As a result, both energy and force evaluations are computed in a single compiled computational kernel, avoiding the combinatorial growth of explicit derivative expressions, leading to substantial savings. Further, higher-order derivatives can also be evaluated within the same framework through repeated application of AD operations, making second- and higher-order derivatives accessible in a consistent way without additional symbolic reformulation. In previous Fortran-based implementations, on the contrary, analytical gradients may be available, but higher-order derivatives are often computed numerically by finite difference.

The descriptor implementation is decoupled from the model type, allowing both linear PIP fitting and NN-based models (PIP-NN / FI-NN) within a unified interface. This design also allows straightforward extension to other polynomial-based invariant representations by modifying only the aggregation or coefficient layer. For practical use in molecular simulations, JaxPIP models can be exported as static computation graphs via ONNX, enabling deployment through standard inference runtimes and integration with external simulation packages.

The workflow of JaxPIP is summarized in **Figure 1(b)**, including basis generation by MSA-2.0, JaxPIP initialization and JIT compilation, ONNX export, and subsequent interface to external simulation engines. This provides a unified pipeline from descriptor construction to deployment. JaxPIP is designed to be compatible with existing computational chemistry workflows. Through a lightweight C-API wrapper, the exported ONNX PES models can be integrated into standard quantum chemistry and molecular simulation packages, including Gaussian (via external),[42] Polyrate,[43] VENUS,[44] RPMDRate,[45] and Caracal.[46]

# 3. Computational Details

## 3.1 Benchmark Setup

All CPU benchmarks were performed on servers with dual AMD EPYC 9654 and 768 GB of memory. To ensure maximum execution consistency and eliminate OS jitter caused by thread migration or context switching, all CPU processes were pinned to a single physical core using the "`taskset -c 0`" command. GPU benchmarks were performed on workstations with Intel Core i9-13900KF, 64 GB of memory, and an NVIDIA RTX 4090 GPU with 24 GB of VRAM. Throughout all benchmarks, the data type was set to FP64 to ensure the numerical precision required for chemical accuracy, as is standard practice in our production simulations.

We measured the performance of three different workloads. Since all three workflows share the same initial step of generating the primitive PIP basis via MSA-2.0, we do not include this shared step below. The metrics were recorded as follows:

(1) MSA-2.0 (The Fortran Baseline)[18, 19, 33]

  (a) Factorization: Time for performing the monomial factorization.

  (b) Compilation: Time spent on compiling a PIP linear model for execution benchmark with generated Fortran code. The compile command is "`gfortran -O3 -march=native -ffast-math -fdefault-real-8 -fdefault-double-8 -ffree-line-length-none`" to maximize the performance.

  (c) Execution: Time for evaluating both energy and analytical forces.

(2) MOLPIPx[34]

  (a) JIT: Time required for the XLA compiler to trace and compile the PIP linear models in MOLPIPx, starting from MSA-2.0's factorized representation.

  (b) Execution: Time for evaluating both energy and analytical forces.

(3) JaxPIP (This Work)

(a) JIT: Time required for the XLA compiler to trace and compile the PIP linear models in JaxPIP directly from the primitive PIP basis.

(b) Execution (JAX Native): Time for evaluating both energy and analytical forces on CPU or GPU. For GPU benchmarks, an additional dimension by varying the batch size (from $2^0$ to $2^{14}$) was introduced to assess the scaling behavior and throughput potential for massive parallel production workloads, such as in PIMD, RPMD and DMC.

(c) Execution (ONNX): The evaluation time of the exported ONNX model using the ONNX Runtime (ORT), which serves as the production path for MD engines written in C, C++, or Fortran.

These benchmarks are designed to reflect per-configuration energy and force evaluation costs in MD and DMC simulations.

**3.2 Diffusion Monte Carlo**

In DMC simulations, the imaginary-time propagation of an ensemble of walkers provides a route to obtain ground-state quantum properties directly from a given PES. Beyond serving as a benchmark for total energy, DMC can also be used to probe the global smoothness and physical consistency of the PES, as any artificial discontinuity or "hole" in the surface can lead to unstable walker dynamics. In addition, DMC enables the evaluation of zero-point energy (ZPE) and quantum-averaged structural properties, providing a direct link between the potential and observable vibrational properties.

The standard unbiased DMC algorithm employs a discrete weighting scheme.[47] The walkers are dynamically created (birth) or destroyed (death) at each step based on the reference energy, an operation known as branching, as shown in **Eq. (10)**:

$$\begin{cases} P_{\text{birth}} = \exp[-(E_i - E_{\text{ref}})\Delta\tau] - 1 & (E_i < E_{\text{ref}}) \\ P_{\text{death}} = 1 - \exp[-(E_i - E_{\text{ref}})\Delta\tau] - 1 & (E_i > E_{\text{ref}}) \end{cases}, \tag{10}$$

where $E_i$ is the $i$-th walker's energy, $E_{\text{ref}}$ is the reference energy, and $\Delta\tau$ is the step size in imaginary time.

The reference energy is updated based on the live walkers as given in **Eq. (11)**:

$$E_{\text{ref}} = \langle V(\tau) \rangle - \alpha \frac{N_{\text{walkers}}(\tau) - N_{\text{walkers}}(0)}{N_{\text{walkers}}(0)}, \tag{11}$$

where $\tau$ is the imaginary time, $\langle V(\tau)\rangle$ is the average potential energy, $N_{\text{walkers}}(\tau)$ is the number of live walkers at time $\tau$, $N_{\text{walkers}}(0)$ is the initial number of walkers, and $\alpha$ is the damping parameter that controls the number of walkers and reference energy.

Conversely, the continuous weighting scheme maintains a fixed number of walkers, an approach originally formalized to eliminate population control bias[48] and recently adopted in molecular vibrational studies.[49, 50] In this approach, each walker has its weight and the weights are updated at every evolution step according to **Eq. (12)**:

$$w_i(\tau + \Delta\tau) = w_i(\tau)\exp[-(E_i - E_{\text{ref}})\Delta\tau], \tag{12}$$

where $w_i$ is the weight of the $i$-th walker, $w_i(\tau)$ and $w_i(\tau + \Delta\tau)$ are the weights at time step $\tau$ and $(\tau + \Delta\tau)$ in imaginary time.

To prevent weight divergence, where a few walkers accumulate most of the total weight, a resampling method is applied. The effective number of walkers ($N_{\text{eff}}$) is estimated by **Eq. (13)**:

$$N_{\text{eff}} = \frac{(\sum w_i)^2}{\sum w_i^2}, \tag{13}$$

when $N_{\text{eff}}$ drops below a threshold (e.g., 0.5 $N$, $N$ is the number of walkers), the ensemble is resampled. Resampling methods such as the systematic resampling can select walkers with probabilities proportional to their current weights.[51] During this process, walkers with large weights are duplicated, and those with negligible weights are eliminated. After resampling, the coordinates and lineages are updated, and the weights of all surviving walkers are reset to the ensemble average weight ($\langle w\rangle = \sum w_i/N$) to preserve the total statistical weight.

The reference energy is updated to maintain population stability. Analogous to the discrete weighting scheme, the update is based on the total weight of the ensemble, according to **Eq. (14)**:

$$E_{\text{ref}}(\tau) = \langle V\rangle - \alpha\frac{\sum w_i(\tau) - \sum w_i(0)}{\sum w_i(0)}, \tag{14}$$

where $\langle V(\tau)\rangle$ is the weight-averaged potential energy, $\sum w_i(\tau)$ is the total weight at the current time step $\tau$, $\sum w_i(0)$ is the initial total weight (identical to the number of walkers), and other parameters are kept the same as in discrete weighting.

Finally, the imaginary-time average of the reference energy over the DMC simulation after reaching equilibration provides an estimate of the zero-point energy (ZPE), while the sampled walker distribution encodes vibrationally averaged structural information.

### 3.3 Quasi-classical Trajectory

To fully utilize the parallel evaluation capability of JaxPIP, we developed a fully vectorized classical machanics integrator. Different from traditional QCT workflows like VENUS96C[44] that thousands of sequential tasks are submitted with each assigned to a single CPU core, our implementation features data-level parallelism over trajectories. We pack thousands of trajectories into a single batch to run on the GPU simultaneously. This allows efficient evaluation of large ensembles of classical trajectories, enabling direct computation of dynamical observables such as reaction probabilities, state-to-state distributions, and integral cross sections. Note that this scheme is also applicable to path integral based methods such as PIMD and RPMD.

To achieve high performance in JAX, the code must fit the XLA compiler, which prefers static computational graphs and fixed tensor shapes. A practical problem in batching QCT (but not in PIMD or RPMD) is that different trajectories typically require different propagation time. Some trajectories react quickly, while others exhibit long dynamics. If we remove finished trajectories and dynamically shrink the batch size, it will trigger XLA recompilation at almost every step, causing significant overhead and degrading GPU performance.

To resolve this issue, we developed a vectorized integrator based on the standard Velocity Verlet algorithm via `jax.lax.scan` to keep the tensor shape strictly constant.[52] A termination function evaluates the phase space (coordinates and momenta) at each step. If a trajectory satisfies the termination condition, it is flagged by a mask. Its coordinates, momenta, and forces are then frozen. The remaining active trajectories continue to evolve, while terminated trajectories perform dummy steps. By using this masking strategy, all trajectories advance synchronously without changing batch size, thus avoiding thread divergence and maintaining high GPU throughput while preserving statistically independent trajectory evolution.

## 4. Results and Discussion

### 4.1 Computational Efficiency

JaxPIP starts directly from the primitive PIP basis without any factorization, preserving the physical requirements of translational, rotational, and permutational symmetry. Upon loading the basis, the PIP descriptors are constructed as a sequence of dense tensor operations rather than a tree of scalar multiplications, as shown in the right pathway in **Figure 1 (a)**. The detailed math and implementation are provided in **Secs. 2** and **3**. This change in architecture resolves the graph-tracing bottleneck found in early work like MOLPIPx, making the JIT compilation almost instantaneously, even for high-order polynomials.

To evaluate the computational efficiency of JaxPIP, we performed extensive benchmarks covering JIT compilation, execution throughput, and scaling behavior on both CPU and GPU platforms. We constructed a benchmark set including prototypical molecular systems ranging from 3 to 9 atoms as shown in **Table 1**. The evaluated polynomial degrees vary by system size: orders 3–8 for 3–5 atoms, orders 4–7 for 6–8 atoms, and orders 4–6 for 9 atoms. This set covers the entire benchmark dataset used in MOLPIPx, extending to complex, high-dimensional reactive and non-reactive systems detailed in our previous studies: the $H_2S$ dimer ($A_4B_2$),[53] Cl + $CH_4$/$SiH_4$ ($A_4BC$),[54, 55] $CH_3CN$ ($A_3B_2C$),[56] H + $CH_3OH$ ($A_5BC$),[57] F/Cl + $CH_3OH$ ($A_4BCD$),[58, 59] and OH + $CH_3OH$ ($A_5B_2C$)[60] systems. To stress-test the upper limits of JaxPIP, an $A_6B_2C$ system is also included up to the $6^{th}$ order. All CPU benchmarks were performed on an AMD EPYC 9654 server, while GPU benchmarks were offloaded to an NVIDIA RTX 4090.

The most significant advantage of JaxPIP is the elimination of the time-consuming factorization and compilation steps required by the MSA-2.0 and MOLPIPx frameworks. A comprehensive timing comparison, including PIP basis generation, JIT compilation, and energy with force evaluation across different molecular sizes and hardware platforms, is provided in the Supporting Information (**Tables S1–S6**). We first focus in **Figure 2(a)** on small-to-medium systems (3–5 atoms) to compare with MOLPIPx. Note that the 5-atom ABCDE system at the 8$^{th}$ order caused an out-of-memory (OOM) error in MOLPIPx during initialization, resulting in the missing data point in the figure. The $A_5$ system at the 8$^{th}$ order could initialize, but it required about 30 minutes (1,783.82 s) for JIT compilation. After JIT, the MOLPIPx evaluation of energy and forces took sub-millisecond to roughly 25 ms per configuration on a single CPU core, as shown in **Figure 2 (b)**.

Conversely, the JaxPIP benchmarks (**Table S4**) show a very different picture. The JIT latency for all these systems is below 1.0 seconds (**Figure 2(a)**), orders of magnitude faster than MOLPIPx. In addition to the instantaneous compilation, the actual evaluation run times of JaxPIP (**Figure 2(b)**) are also significantly shorter than those of MOLPIPx.

A unique feature of JaxPIP is that its performance is almost independent of symmetry. In MSA-2.0 or MOLPIPx, changing the symmetry group for the same number of atoms can lead to huge differences in factorization and JIT times (**Figure 2(a-b)**). In JaxPIP, the workload depends mainly on the number of flat monomials ($\boldsymbol{\tau}$ in **Eq. (7)**). As shown in **Tables S3** and **S4**, systems with the same atom count and order have the same number of flat monomials, regardless of their symmetry.

The performance data for larger systems (6-9 atoms) are shown in **Figure 3**, which tells the same story. The preparation in MSA-2.0 takes a very long time. For a production-scale 8-atom $A_5B_2C$ system at the 5$^{th}$ order, the monomial factorization took 3,264.89 seconds (~54 minutes) and Fortran compilation took another 33 minutes. This total wait of 1.5 hours must be repeated every time when trying different orders of PIPs or migrating the code to other servers. The cost rises sharply with complexity: for the same $A_5B_2C$ system at 6$^{th}$ order, MSA-2.0 requires 4.5 days for factorization. By bypassing factorization, JaxPIP completes JIT in only 3.58 seconds for the 5th-order $A_5B_2C$ system. This is a $10^5$-fold speedup in the preparation phase compared to MSA-2.0. It is also worth noting that MOLPIPx is unable to handle systems of this scale at all, as it fails during the initialization or JIT stages for these larger cases.

On a single CPU core, the factorized Fortran kernel from MSA-2.0 is faster, evaluating the linear PIP model of $A_5B_2C$ system in 1.42 ms, evaluating both energy and forces, compared to 8.51 ms for JaxPIP. When exported to ONNX format, JaxPIP takes 9.53 ms using the standard ONNX Runtime. While the Fortran code is faster, its performance depends heavily on high compiler optimization (like `-O3`), which further slows down the compilation. On the other hand, JaxPIP's speed is already "production-ready" for standard MD simulations. More importantly, JaxPIP is much more flexible: unlike MSA-2.0, which needs re-compilation for every new system, we can switch between different molecules simply by loading different ONNX files.

Unlike Fortran kernels built for single-core CPU speed, JaxPIP is designed for GPU acceleration from the start. Its tensorized architecture is perfect for methods like PIMD, RPMD, and DMC, which require evaluating thousands of beads or walkers at once. In these benchmarks, we focus on the amortized evaluation time (total batch time divided by batch size). Since hundreds or thousands of configurations must move forward together in quantum simulations, our goal is to finish the entire batch as fast as possible.

**Figure 4** shows that moving to GPU does not increase JIT latency. For example, JIT for the $A_5B_2C$ system takes only about 4 seconds on the RTX 4090. **Figure 5(a-f)** shows how throughput improves with batch size on 6-9 atom systems. The missing data points in the upper-right regions represent the OOM limits. It is worth noting that performance stays very stable right up to this hardware limit, with only a slight increase in per-sample latency just before an OOM occurs. For almost all systems, the cost per sample on GPU drops below CPU cost once the batch size exceeds $2^3$ (8 configurations). As the batch size grows, the per-sample time hits a plateau ($10^{-3}$ to $10^{-2}$ ms), suggesting the GPU is fully saturated. On a 24 GB RTX 4090, we can fit a batch size of up to $2^{12}$ (4,096 configurations) for the $A_5B_2C$ system before hitting memory limits. This capacity is more than enough for typical PIMD/RPMD (32–128 beads) or DMC (thousands of walkers) calcuations, indicating that JaxPIP can handle complex quantum dynamics on a gaming GPU.

### 4.2 DMC Simulation of Glycine

Diffusion Monte Carlo (DMC) is a powerful approach for vibrational problems in molecular systems, specifically for determining the ZPE and ground-state wavefunction. As outlined in **Sec. 2**, this method evaluates the exact ground state by propagating the time-dependent Schrödinger equation in imaginary time using an ensemble of walkers.[47, 61-63]

We performed DMC simulations of all eight conformers of glycine to calculate accurate ZPEs, using the full-dimensional fragment PIP PES developed by Conte et al.[64] Identical simulation parameters were used as reported in a previous DMC study: 30,000 walkers, a step size of $\Delta\tau$ = 1.0 a.u., and 55,000 steps (15,000 steps for equilibration and 40,000 steps for production).[65] The primary methodological distinction in our work is the application of the continuous weighting scheme, which enhances computational performance on GPU architectures. To estimate the statistical uncertainty of the ZPEs, three independent simulations were performed for each conformer.

In our simulations, JaxPIP directly loaded the PES weights from the original fragment, ensuring the same PES is used. **Figure 6** compares our results with literature values obtained from the fragment PIP[64] and the ANI PES.[65] The absolute differences between ZPEs produced by JaxPIP and the original fragment PIP are all within 20 cm$^{-1}$ (~0.057 kcal/mol), which we can attribute to the difference in the DMC algorithms (continuous vs. discrete weighting). Our simulations successfully reproduced the expected ZPE features, where specific conformers share nearly identical ZPEs, classified into four pairs. The calculated ZPE differences between the pairs are minimal: 6 cm$^{-1}$ for conformers 1 & 4, 4 cm$^{-1}$ for conformers 3 & 5, 2 cm$^{-1}$ for conformers 2 & 7, and 1 cm$^{-1}$ for conformers 6 & 8.

Regarding computational efficiency, a single DMC simulation including 30,000 walkers and 55,000 steps took 9.5 hours on an NVIDIA RTX 4090 GPU using JaxPIP. To fit into the VRAM, the 30,000 walkers were split into 3 chunks, and the potential energy of 10,000 walkers was evaluated sequentially in each step. We further validated multi-GPU scaling on a machine equipped with an AMD EPYC 7302P CPU and three

RTX 4080 Super GPUs (16 GB VRAM each). By distributing the walkers across the three GPUs, the total wall time was reduced to 6.3 hours (approximately 2/3 of the single RTX 4090 runtime).

Compared with existing implementations, JaxPIP demonstrates significant efficiency gains: it is 3.1 times faster than the original fragment PIP (30 hours on a Xeon CPU)[64] and 6.3 times faster than the ANI model (60 hours on an RTX 3090 GPU).[65] Given that FP64 precision is required for chemical accuracy, we expect even higher performance on data-center GPUs such as NVIDIA A100 or H100, which feature native high-throughput FP64 units and larger high-bandwidth memory.

**4.3 Quasi-classical Trajectory**

To demonstrate the applicability of JaxPIP to realistic reaction dynamics, we perform QCT calculations for the OH + $CH_3OH$ system. It is a prototypical dual-channel hydrogen abstraction reaction widely studied in gas-phase reaction dynamics.[60, 66] In this work, our previously developed FI-NN PES[66] was migrated to the JaxPIP framework. The model consists of 419 input FIs, two hidden layers with 20 and 80 neurons, respectively, and a single-output architecture, resulting in 10,161 parameters. Numerical consistency between the original implementation and the JaxPIP version is tested and confirmed, with energy differences remaining within standard FP64 precision numerical accuracy.

To compare the numerical consistency and computational performance of JaxPIP in practical reactive dynamics, QCT simulations were performed using both the original Fortran-based workflow and the JAX-native implementation. Specifically, calculations were carried out using the original Fortran FI-NN PES with VENUS96C, as well as the migrated FI-NN PES evaluated through JaxPIP within our vectorized JAX-native QCT engine. Depending on the collision energy, 2–7 × $10^4$ trajectories were propagated at collision energies of 1, 2, 5, 8, 10, 15, and 20 kcal/mol. At the beginning of each trajectory, the two reactants were separated by 10 Å. For both implementations, the reactants were initialized in their ro-vibrational ground states. Initial conditions were generated using the standard VENUS96C and VENUSPy, respectively.[67] The detailed impact parameter ($b_{max}$), trajectory statistics, and reactive events for both methyl (R1) and hydroxyl (R2) hydrogen abstraction channels are summarized in **Table S7**. **Figure 7** compares the resulting integral cross sections (ICSs) obtained from the original Fortran implementation and JaxPIP. Across the full collision-energy range, the two implementations show nearly indistinguishable results, confirming that migration to the tensorized JaxPIP framework preserves the dynamical behavior of the original PES. The minor discrepancies between the two sets of QCT results are primarily attributed to the statistical uncertainties inherent in the finite number of trajectories. Furthermore, differences between the velocity Verlet integrator utilized by the JAX-native engine and the combined fourth-order Runge-Kutta and sixth-order Adams-Moulton predictor-corrector integrator utilized in VENUS96C,[44] together with the sampling of initial conditions, also contribute marginally. As presented in **Table S8**, good agreement between the Fortran FI-NN and JaxPIP is achieved for channel specific ICSs, total ICSs, and the product branching ratios across all collision energies.

Having established numerical equivalence, we next examine the computational characteristics of the JAX-native implementation. Unlike traditional VENUS96C workflows, where trajectories are distributed across independent CPU cores, the tensorized structure of JaxPIP enables large batches of trajectories to be propagated simultaneously on GPU hardware. **Figure 8** summarizes the scaling behavior with respect to batch size and demonstrates the throughput advantage of vectorized trajectory propagation. The batch-size scaling behavior of the FI-NN model exhibits a different pattern compared to the linear PIP fitting discussed in **Sec. 4.1**. For the PIP linear fitting, the amortized time per sample generally decreases and hits a flat plateau until an OOM error occurs, as shown in **Figure 5**. However, the FI-NN model displays a clear sweet spot at a smaller batch size, which is significant for choosing an optimal number of trajectories to maximize the overall computational throughput on the GPU. The run time per sample reaches its minimum at a batch size of $2^9$ (512 configurations), taking roughly ~$1.0 \times 10^{-5}$ seconds per sample. When the batch size further increases from $2^{10}$ to $2^{12}$, the time per sample goes up rather than down.

To provide a practical perspective on this throughput advantage, we benchmarked a representative ensemble of 20,480 trajectories at the collision energy of 20 kcal/mol. Running the Fortran-based VENUS96C workflow parallelized across an entire dual AMD EPYC 9654 CPU server (192 cores, 384 threads) required approximately 9 minutes of wall time, amounting to 22 core-hours (or an amortized CPU time of 3.8 s per trajectory). In contrast, JaxPIP processes trajectories in batches. As detailed in **Sec. 3.3**, our masking strategy requires all trajectories within a batch to perform dummy steps until the longest-lived trajectory finishes. Despite the computational overhead introduced by these dummy steps, distributing the same 20,480 trajectories across a supercomputing cluster with 320 NVIDIA V100 GPUs (split into 320 tasks of 64 trajectories each) remarkably reduced the total wall time to less than 1 minute (with the longest task taking 49 s). This corresponds to a total of 3 card-hours, or an amortized GPU time of 0.5 s per trajectory per card. Crucially, this demonstrates a 7-fold reduction in device-hours (3 card-hours vs. 22 core-hours) and per-trajectory evaluation cost (0.5 s vs. 3.8 s), alongside an 11-fold wall-clock speedup compared to the fully parallelized 192-core CPU server.

### 4.4 Discussion

To understand the design in JaxPIP, it is useful to revisit the historical trade-offs in potential evaluation. For an $N$-atom system, there are $N(N-1)/2$ pairwise internuclear distances. Traditional factorization strategies, such as MSA-2.0, are designed to reduce the number of expensive transcendental functions, in particular the exponential function. Morse-like intermediate variables are evaluated once and reused, and the remaining polynomial expansion is constructed through a large number of scalar multiplications. For small systems (e.g., a 5-atom case), this corresponds to a small number of exp evaluations followed by a deep multiplication tree.

Over the past decade, improvements in arithmetic throughput in modern CPU and GPU architectures have outpaced the growth of memory bandwidth. From the perspective of the Roofline model,[68, 69] this shift implies that many scientific workloads are increasingly constrained by memory access and data movement, rather than floating-point arithmetic. This effect is particularly pronounced for workloads with irregular memory access or low arithmetic intensity, where reducing memory traffic can yield larger performance gains than reducing FLOPs.[68, 69]

Historically, such space-for-time trade-offs were particularly effective on earlier CPU architectures with limited vector width and fewer execution units, where reducing total FLOP count often correlated well with reduced runtime for compute-dominated kernels. However, even in these systems, memory hierarchy effects (e.g., cache capacity and bandwidth) already played an important role. For small molecular systems, intermediate data structure may fit within higher levels of cache, making memory latency less critical, but this balance changes as system size and model complexity increase.

In MSA's factorization approach, higher-order polynomials (**p**) are recursively generated from a pool of pre-computed lower-order monomials (**m**). As both the system size and polynomial order increase, the number of intermediate monomials grows much faster than the number of final polynomials. As shown in **Table S1** and **S2** for representative molecular systems (3-9 atoms), this gap becomes increasingly pronounced at higher orders. For example, in the $A_5$ system, the 3$^{rd}$ order basis contains 176 monomials used to construct 12 polynomials, while the 8$^{th}$ order expansion requires 10,158 monomials for 580 polynomials.

This scaling behavior directly affects gradient-based optimization and molecular dynamics, where derivatives with respect to atomic coordinates are required. The gradient is computed using the chain rule, $\mathbf{J}_{p(xyz)} = \mathbf{J}_{p(m)}\mathbf{J}_{m(xyz)}$, where $\mathbf{J}_{p(m)}$ is the Jacobian of the polynomials (**p**) with respect to the monomials (**m**), and $\mathbf{J}_{m(xyz)}$ is the Jacobian of the monomials (**m**) with respect to coordinates (*xyz*). When both the number of polynomials and monomials becomes large, the intermediate Jacobian terms can no longer be handled efficiently without affecting memory usage. In practice, this leads to large intermediate data structure during evaluation, which increases memory traffic. In our previous work on the OH + $CH_3OH$ ($A_5B_2C$) system, for example, the 5$^{th}$-order PIP basis contains 3,060 final polynomials but 60,460 intermediate monomials.[60] Although these monomials are mathematically eliminated in the final contraction, they may still need to be constructed or stored during evaluation depending on the implementation. In double precision (FP64), a dense storage of this intermediate representation corresponds to about 11.6 MB per molecular configuration.

In practical simulations such as QCT,[44] PIMD,[70] RPMD,[45, 46] and DMC,[71] a common parallelization strategy is to assign one trajectory to one CPU core. In this regime, each core operates with its own local workload, and performance is closely related to the cost of evaluating the energy and forces for a single molecular configuration. This motivates reformulating the evaluation to avoid explicit construction of large intermediate representations.

The compilation behavior of MOLPIPx can be understood by contrasting traditional symbolic compilers with modern AD systems. Modern AD frameworks such as XLA operate by tracing the program and constructing an intermediate computational graph before execution. In MOLPIPx, the evaluation still relies on a deeply nested structure generated by MSA. As a result, the AD tracer must expand this dependency structure during graph construction. Since the computation is dominated by many small scalar operations rather than large tensor operations, there is limited opportunity for graph fusion or kernel-level optimization. This leads to a large and fragmented computational graph, which increases compilation overhead and JIT latency, suggesting that the inefficiency arises not only from the number of operations, but also from the highly fragmented structure of the evaluation.

JaxPIP addresses these issues by changing the representation of the evaluation rather than reducing the number of operations. Instead of directly evaluating the monomial product trees, the computation is reformulated using a logarithmic transformation, $\prod y_k^{b_k} = \exp[\sum b_k \ln y_k]$. This transformation is not introduced for numerical convenience, but to reshape the evaluation into a form that is more suitable for modern execution models. In traditional implementations, direct evaluation of algebraic products is usually preferred to avoid additional transcendental operations such as exp and log. However, once the polynomial structure is fixed, the evaluation can be rewritten in terms of linear algebraic operations on a fixed coefficient matrix.

By mapping the sparse dependencies into the logarithmic domain, the evaluation becomes dense and highly regular. In this formulation, the exponent matrix **B** has dimensions $N_{\mu} \times N_{\mathrm{d}}$. For the $A_5B_2C$ system at 5$^{th}$ order, this corresponds to a matrix of size 237,336 × 28, or about 51 MB in FP64, but only about 6 MB in Uint8. This representation removes the irregular structure of the original evaluation and reduces memory usage to ~1/8 without losing any precision. As a result, the computation becomes more uniform and better aligned with optimized numerical routines for dense linear algebra.

In summary, JaxPIP trades an increase in arithmetic operations for a more regular and structured computation. While this may appear less efficient from a FLOPs perspective, it improves the consistency of the evaluation by reducing structural irregularities. This is particularly relevant for molecular simulations such as QCT, PIMD, RPMD, and DMC, where the accurate and efficient evaluation of potential energies and forces for a large number of nuclear configurations directly determines the convergence of chemically meaningful observables.

## 5. Conclusions and Outlook

It is important to emphasize that PIP-based methods have played a central role in making high-fidelity PES modeling a routine tool for molecular dynamics and kinetics in gas-phase systems. The monomial symmetrization approach (MSA) enabled evaluation of high-order PIP descriptors through recursive factorization and common sub-expression reuse and also supported the development of automated analytic gradient generation. MOLPIPx extended this workflow into the JAX framework, allowing gradients to be obtained via automatic differentiation without explicit symbolic derivation, although at the cost of long JIT compilation times for large systems. In this work, JaxPIP resolves these limitations by reformulating PIP and FI evaluation into a tensorized JAX-native architecture that enables efficient GPU execution with scalable automatic differentiation. In addition, it naturally supports higher-order derivatives through automatic differentiation, enabling efficient access to Hessians as well as third- and fourth-order derivative tensors. This provides a unified framework for evaluating energies, forces, and response properties within the same computational graph.

This design directly translates into practical efficiency gains for chemical dynamics simulations. For example, in QCT calculations of the OH + $CH_3OH$ system, $2\times10^4$ trajectories per collision energy can be propagated across a wide energy range within approximately 33 minutes on a single NVIDIA RTX 4090 GPU using FP64 precision. At the PES model level, the migrated JaxPIP implementation is numerically consistent with the original Fortran reference; at the dynamical level, the resulting trajectory ensembles reproduce statistical observables, with no deviation in converged reaction probabilities, ICSs, and branching ratios compared to the baseline implementation. Similar performance characteristics also extend to quantum nuclear simulations such as DMC and vibrationally averaged properties, where large ensembles of walkers or trajectories can be evaluated in a fully vectorized manner without CPU-level parallelization overhead.

These results indicate that reformulating PIP (or FI) evaluation as dense tensor operations enables GPU-native execution with orders-of-magnitude speedup while preserving machine-precision consistency in both energies, forces, and ensemble-level dynamical observables. This provides a practical receipe for running statistically converged QCT and DMC calculations for medium-sized reactive systems within manageable computational budgets. JaxPIP is being integrated as a PES evaluation engine for trajectory and quantum dynamics simulations in the CQPES package.[72]

**Supplementary Information**

The Supporting Information is available free of charge at [link].

Comprehensive benchmark results, including wall-clock times for PIP basis generation, JIT compilation, and simultaneous energy-force evaluations across various molecular systems (3- to 9-atom) and hardware architectures (AMD EPYC 9654 CPU and NVIDIA RTX 4090 GPU) using MSA-2.0, MOLPIPx, and JaxPIP (**Tables S1–S6**). Detailed QCT statistics and reactive events for the OH + $CH_3OH$ reaction using JaxPIP (**Table S7**) and comparison of ICSs and product branching ratios between the Fortran FI-NN and the migrated JaxPIP PES across all collision energies (**Table S8**).

**Author Contributions**

[#]J.L. (Junhong Li) and K.S. contributed equally to this work.

**J.L. (Junhong Li):** Conceptualization (equal); Methodology (lead); Software (lead); Investigation (equal); Formal analysis (equal); Data curation (lead); Writing – original draft (equal); Writing – review & editing (equal).

**K.S.:** Conceptualization (equal); Investigation (equal); Formal analysis (equal); Writing – original draft (equal); Writing – review & editing (equal).

**H. G.**: Conceptualization (equal); Supervision (equal), Writing – review & editing (equal).

**J.L. (Jun Li):** Conceptualization (equal); Supervision (lead); Writing – review & editing (equal).

All authors have given approval to the final version of the manuscript.

**Data Availability**

The JaxPIP software is available at https://github.com/CQPES/JaxPIP. The specific Permutation Invariant Polynomial (PIP) and Fundamental Invariant (FI) basis sets, along with the benchmark data used in this work, are hosted at https://github.com/CQPES/JaxPIP-Basis-Library. JaxPIP is integrated into CQPES, which is available at https://github.com/CQPES/cqpes-legacy. All links were accessed on Auguest 12, 2026. All source code and data are released under the BSD 2-Clause "Simplified" License.

**Acknowledgements**

This work was supported by the National Natural Science Foundation of China (Grant Nos. 22473019 and W2512014). The authors thank the Open Source Supercomputing Center of S-A-I for providing computational resources. The authors thank the current and former group members for providing the raw data and the PESs for various systems: Feiwen Deng ($H_2S$ dimer), Dr. Yang Liu ($Cl + CH_4$), Xiaohu Xu ($Cl + SiH_4$), Junlong Li ($CH_3CN$), and Dr. Dandan Lu ($F/Cl/H + CH_3OH$). The authors thank Bo Li for the insightful discussions on the JAX implementation, and Hui Ni for valuable discussions on JAX model evaluation via ONNX.

**Conflicts of Interest**

The authors declare no competing financial interest.

**Table 1.** Molecular systems included in the benchmark sets, categorized by the number of atoms.

| # of atoms | Molecules |
| --- | --- |
| 3 | $A_3$, $A_2B$, ABC |
| 4 | $A_4$, $A_3B$, $A_2B_2$, $A_2BC$, ABCD |
| 5 | $A_5$, $A_4B$, $A_3B_2$, $A_3BC$, $A_2B_2C$, $A_2BCD$, ABCDE |
| 6 | $A_4B_2$, $A_4BC$, $A_3B_2C$ |
| 7 | $A_5BC$, $A_4BCD$ |
| 8 | $A_5B_2C$ |
| 9 | $A_6B_2C$ |

**Figure 1. (a)** Comparison between MSA's factorization approach and JaxPIP's tensorization approach; **(b)** Workflow of JaxPIP from basis generation to simulation.

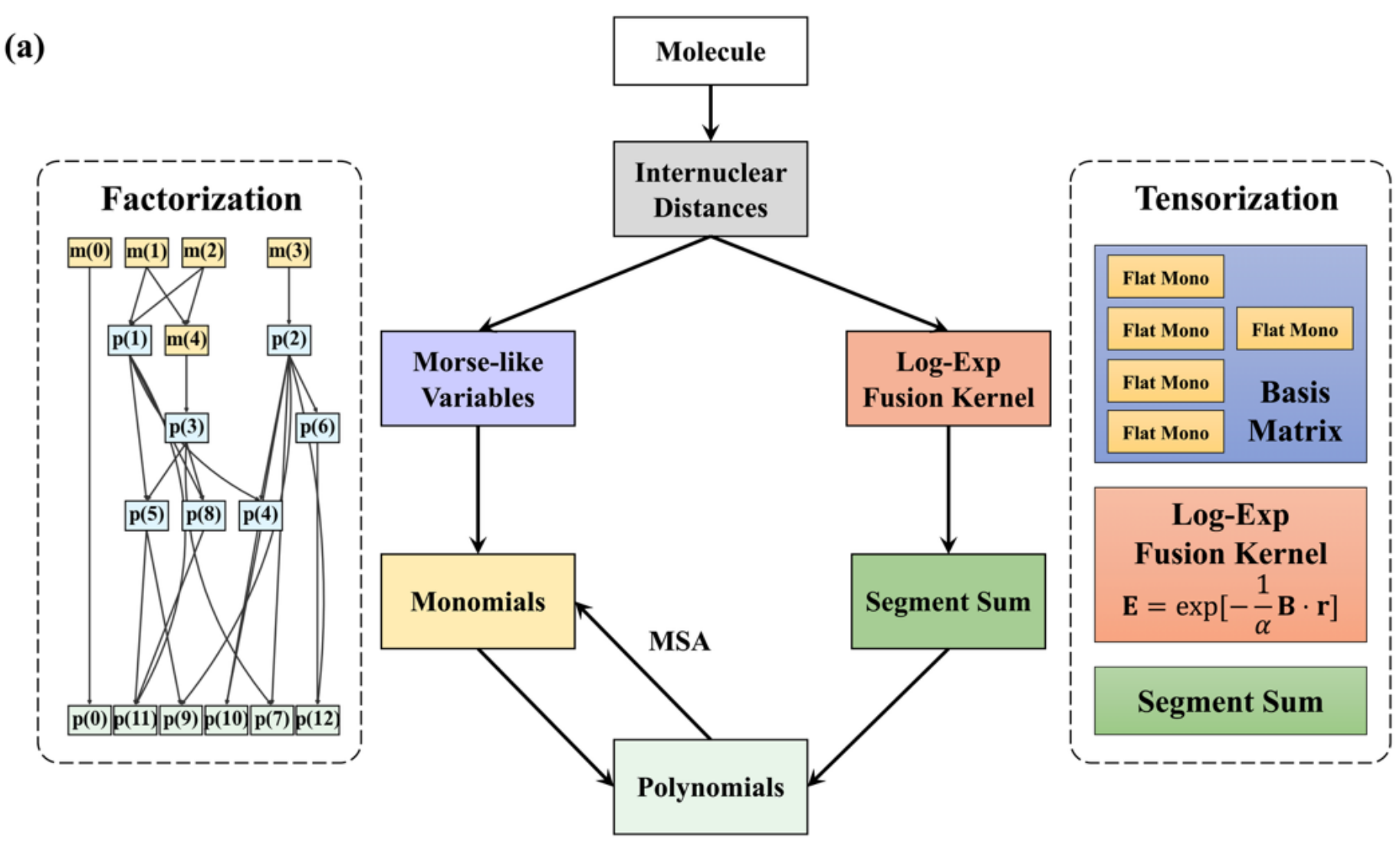

(b)

Permutation Invariant Polynomial (PIP)

Polynomial Basis

Fundamental Invariant (FI)

JaxPIP

ONNX Model

Simulation Engines (C++/Fortran)

Simulation Engines (Python Native)

**Figure 2.** Performance and complexity benchmarks across 3- to 5-atom systems (polynomial orders 3–8) using MOLPIPx (square markers, solid lines) and JaxPIP (cross markers, dashed lines). Vertical dashed lines categorize the benchmarks by total atom count. Timings were measured on a single core of a dual AMD EPYC 9654 CPU server (768 GB RAM) using 64-bit floating-point (FP64) arithmetic.

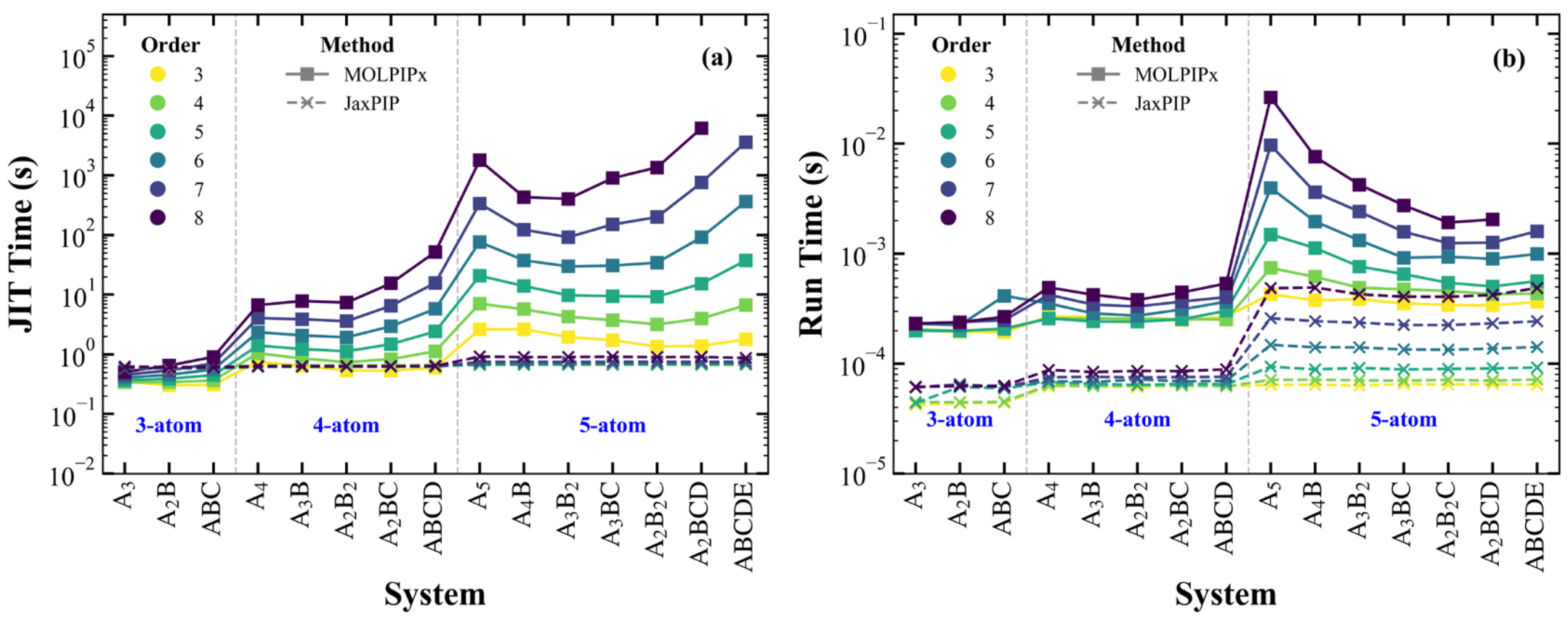

**Figure 3.** Performance and complexity benchmarks across 6- to 9-atom systems (orders 4–7) using JaxPIP. **(a)** JIT compilation times and **(b)** run times for simultaneous energy and force evaluations. Timings were measured on a single core of a dual AMD EPYC 9654 CPU server (768 GB RAM) using 64-bit floating-point (FP64) arithmetic.

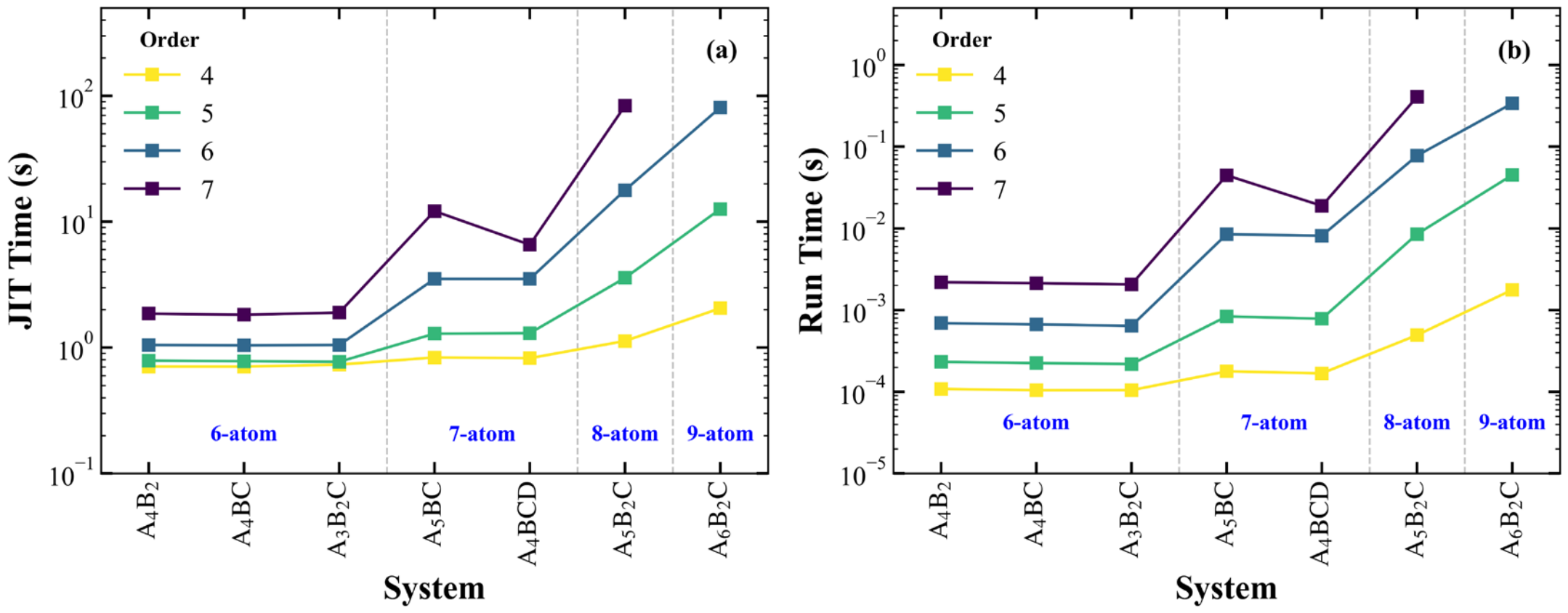

**Figure 4.** JIT compilation overhead and scaling analysis across 6- to 9-atom systems (orders 4–7). Vertical dashed lines categorize the benchmarks by total atom count. Timings were measured on an NVIDIA RTX 4090 GPU (24 GB VRAM) using 64-bit floating-point (FP64) arithmetic.

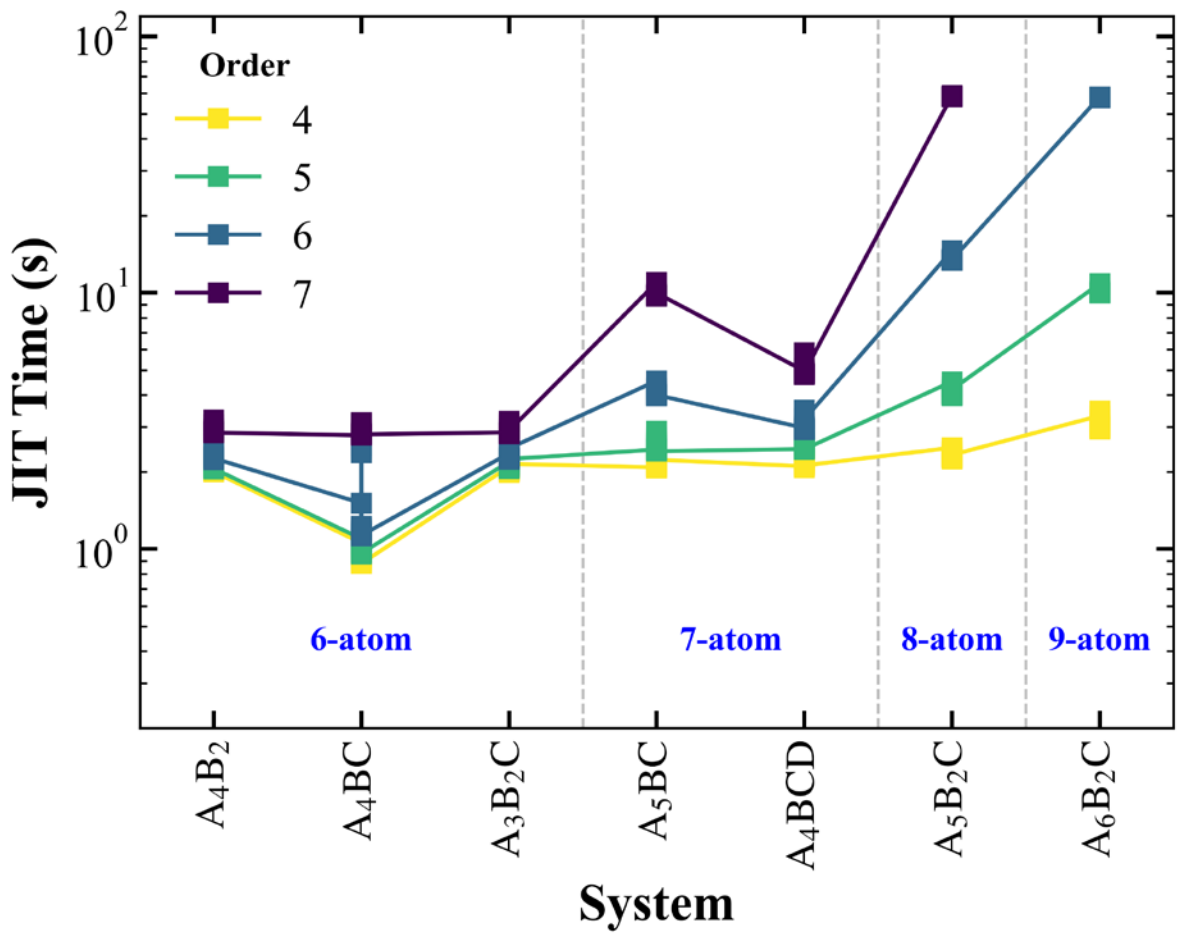

**Figure 5.** GPU batch-size scaling and throughput analysis across 6- to 9-atom systems (orders 4–7) using JaxPIP. Panels **(a)**–**(f)** illustrate the mean evaluation time per sample as a function of batch size for linear PIP models. Timings were measured on an NVIDIA RTX 4090 GPU (24 GB VRAM) using 64-bit floating-point (FP64) arithmetic. The absence of data points in the top-right regions is simply due to out-of-memory (OOM) occurred.

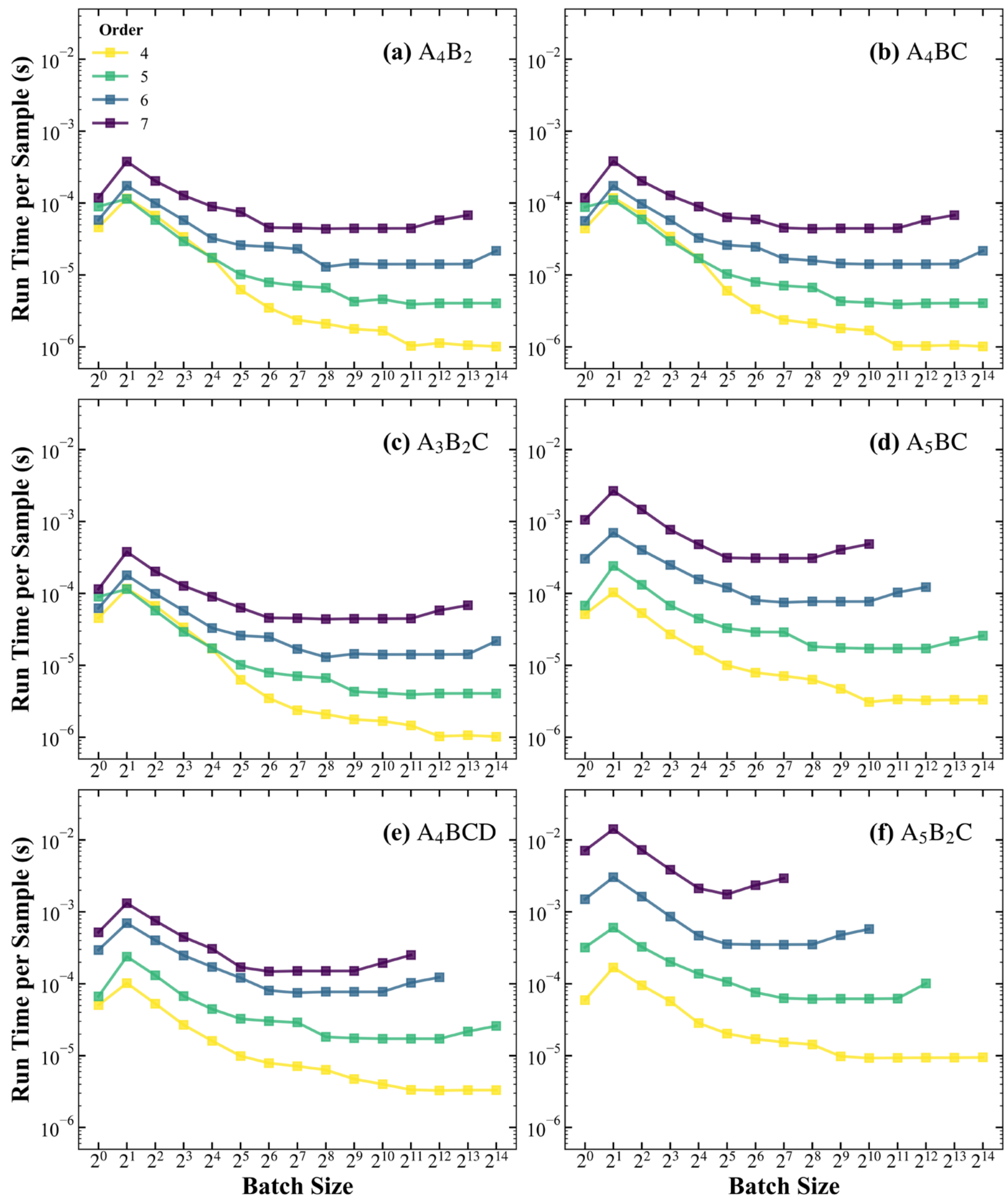

**Figure 6.** Comparison of zero-point energies of glycine conformers 1-8, in $cm^{-1}$. JaxPIP results (black, this work) were obtained using our implementation of the continuous weighting DMC scheme. The PIP (red) and best ANI (blue) results were obtained from refs. [64] and [65].

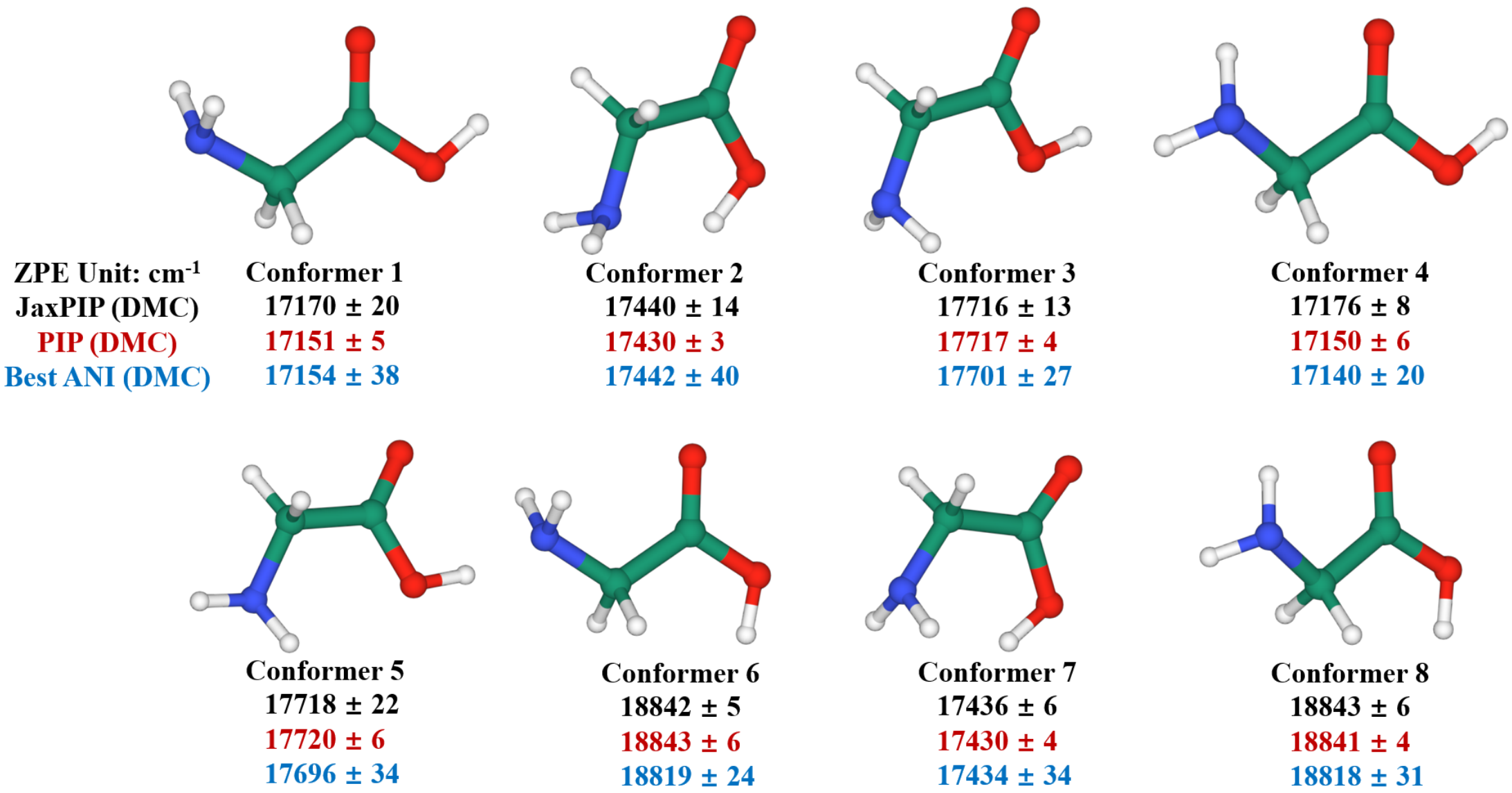

**Figure 7.** ICS results obtained from the Fortran FI-NN PES (green line with dot markers) and the migrated JaxPIP PES (red dashed line with cross markers) as a function of collision energy (in kcal/mol) for the reaction OH + $CH_3OH$. R1 for methyl hydrogen abstraction ($H_2O$ + $CH_2OH$) and R2 for hydroxyl hydrogen abstraction ($H_2O$ + $CH_3O$).

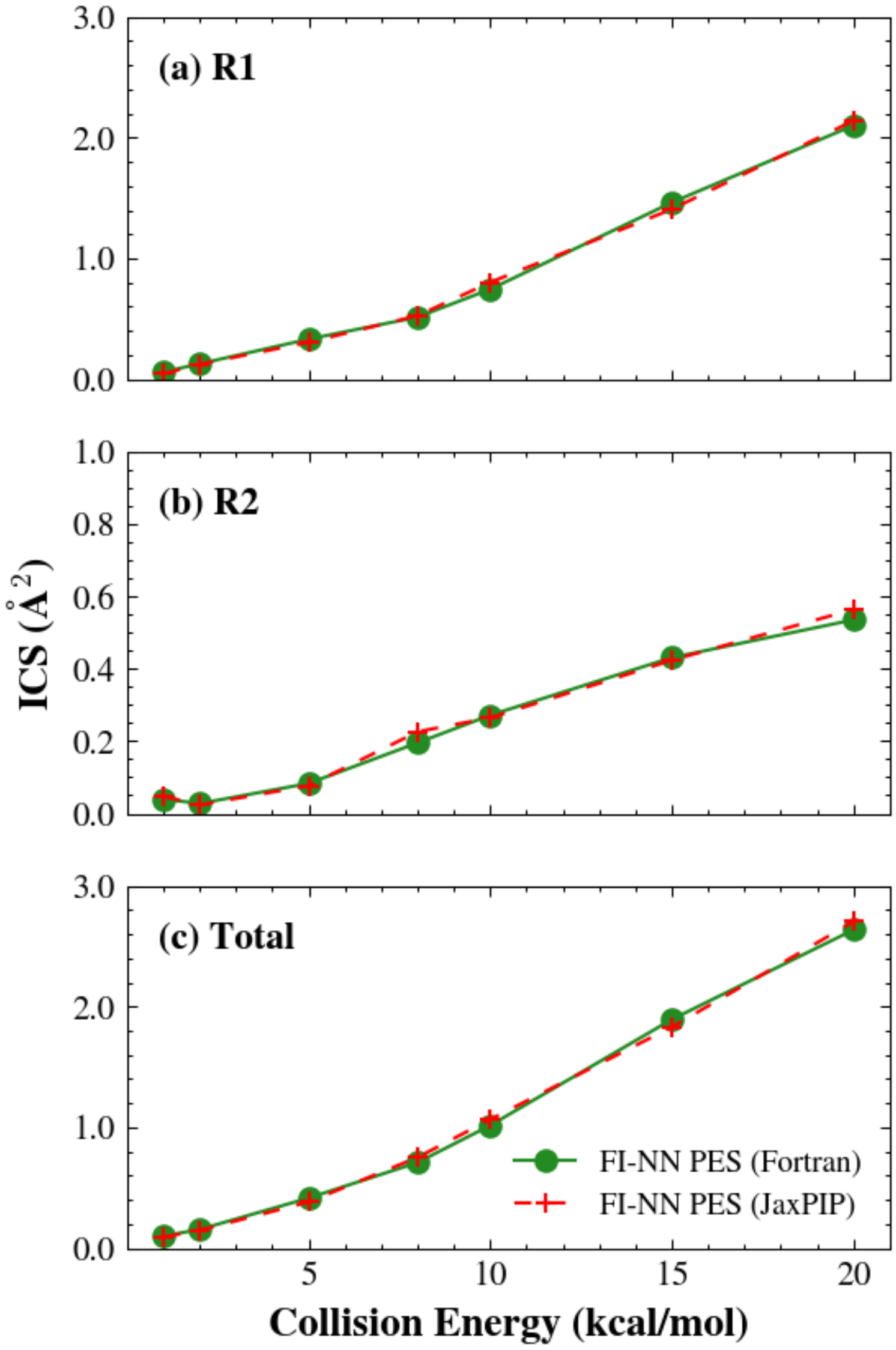

**Figure 8.** GPU batch-size scaling and throughput of the FI-NN model for the OH + $CH_3OH$ system using JaxPIP. Amortized time per trajectory is reported including both energy and force evaluations. The calculation was performed on an NVIDIA RTX 4090 GPU (24 GB VRAM) in FP64 precision.

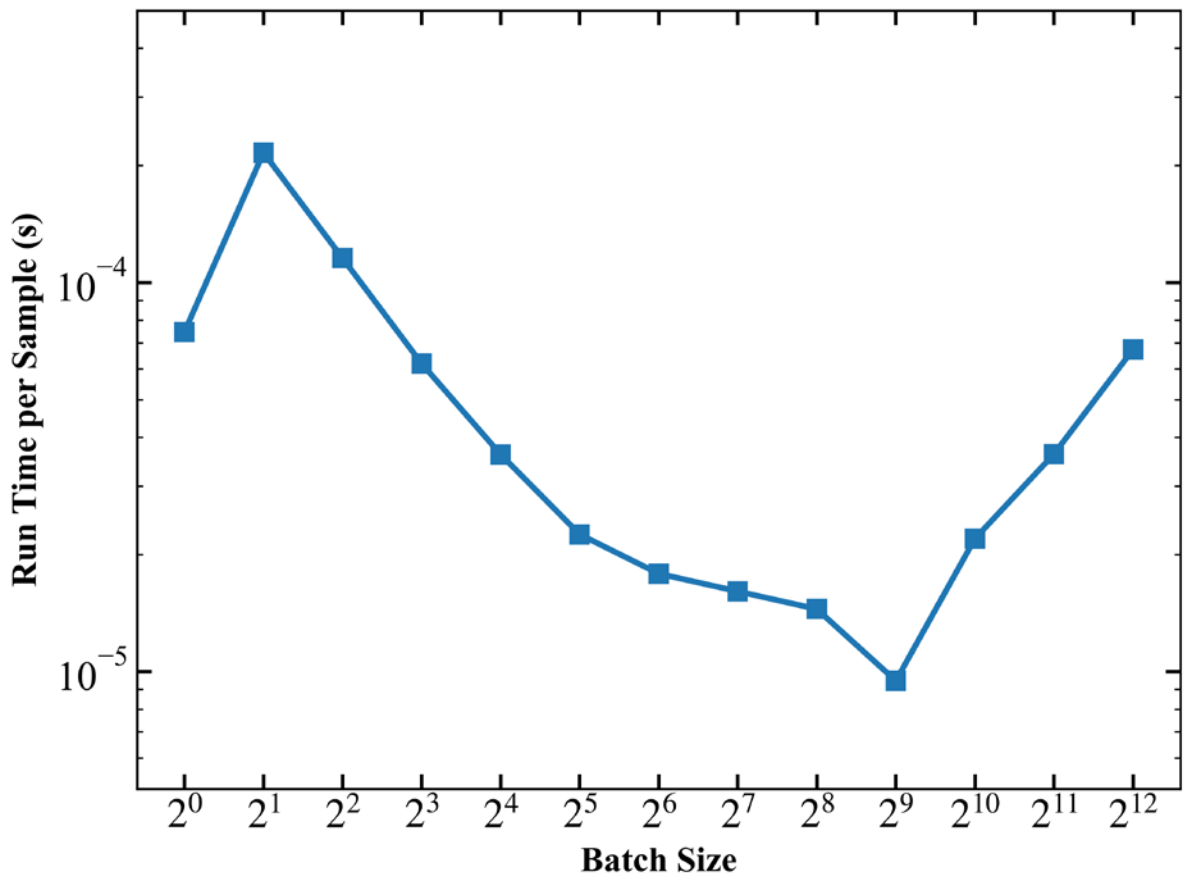